\documentclass{article}

\usepackage{arxiv}

\usepackage[utf8]{inputenc} % allow utf-8 input
\usepackage[T1]{fontenc}    % use 8-bit T1 fonts
\usepackage{hyperref}       % hyperlinks
\usepackage{url}            % simple URL typesetting
\usepackage{booktabs}       % professional-quality tables
\usepackage{amsfonts}       % blackboard math symbols
\usepackage{nicefrac}       % compact symbols for 1/2, etc.
\usepackage{microtype}      % microtypography
\usepackage{graphicx}
\usepackage[caption=false,font=footnotesize]{subfig}
\usepackage{cite}
\usepackage{doi}
\usepackage{amsmath}
\usepackage{textcomp}
\usepackage{multicol}
\usepackage{algorithm}
\usepackage{algorithmic}

\title{Multi-Agent Deep Deterministic Policy Gradient Algorithm for Peer-to-Peer Energy Trading
	Considering Distribution Network Constraints}
\author{{\hspace{1mm}Cephas~Samende}\thanks{All the authors are with the School of Computing and Mathematics, Keele University} \\
	Keele University\\
	United Kingdom 
	\And
	{\hspace{1mm}Jun~Cao}\\
	Keele University\\
	United Kingdom 
	\And
	{\hspace{1mm}Zhong~Fan}\\
	Keele University\\
	United Kingdom \\

}

\date{}

\begin{document}
\maketitle

\begin{abstract}
In this paper, we investigate an energy cost minimization problem for prosumers participating in peer-to-peer energy trading. Due to (i) uncertainties caused by renewable energy generation and consumption, (ii) difficulties in developing an accurate and efficient energy trading model, and (iii) the need to satisfy distribution network constraints, it is challenging for prosumers to obtain optimal energy trading decisions that minimize their individual energy costs. To address the challenge, we first formulate the above problem as a Markov decision process and propose a multi-agent deep deterministic policy gradient algorithm to learn optimal energy trading decisions. To satisfy the distribution network constraints, we propose distribution network tariffs which we incorporate in the algorithm as incentives to incentivize energy trading decisions that help to satisfy the constraints and penalize the decisions that violate them. The proposed algorithm is model-free and allows the agents to learn the optimal energy trading decisions without having prior information about other agents in the network. Simulation results based on real-world datasets show the effectiveness and robustness of the proposed algorithm.
\end{abstract}

% keywords can be removed

\section{Introduction}
Peer-to-peer (P2P) energy trading is a promising approach for addressing the world's \textquoteleft energy trilemma\textquoteright (i.e. environmental sustainability, energy equity, and energy security) facing human society today \cite{zhou2020state}. Its  emergence is as a result of rapid deployment and  connectivity of distributed energy resources (DERs)  to the power system \cite{tushar2020peer}. Conventionally, power systems were dominated by centralized generators situated in strategic locations \cite{wood2013power}. The generated power was transmitted over long distances to consumers for consumption.  As the result, power flow was unidirectional (flowing from generators to consumers) and control of the power flow was easy due to centralized structures (i.e. generation, transmission, distribution and consumption) \cite{wood2013power}. 

With digitization and the emergence of DERs such as battery energy storage systems, rooftop solar photovoltaic (PV) installations and smart home appliances, power systems are no longer passive but active with DERs actively involved in the electricity system \cite{7949037, tushar2020peer}. As DERs can generate energy at point of consumption, power flow in today's power system is bidirectional, posing significant challenges in terms of planning, operation, control and protection of the power system \cite{tushar2020peer, camarinha2016collaborative}. At the same time, the emergence of DERs along with digitization have created new opportunities which can be used to solve most of the challenges caused by DERs through  development of local energy markets such as P2P energy trading schemes \cite{abrishambaf2019towards}.

With P2P energy trading, customers with DERs (called \textquoteleft prosumers\textquoteright as they are able to generate and consume energy) can locally trade and share energy with each other. As many DERs are stochastic in nature, any surplus energy can be sold to neighbouring prosumers with deficit energy via a cloud-based P2P platform. The main role of the P2P platform is to set the P2P energy selling and buying price. 

To encourage prosumer participation in P2P energy trading, the selling and buying price must be higher and lower than the export and import prices imposed by the energy service provider (ESP) respectively \cite{long2018peer}. Thus, prosumers acting as energy producers benefit more from  individual profits and prosumers acting as energy consumers benefit more from cheap energy when they trade with each other via the P2P platform than when they trade directly with the ESP. This creates a win-win situation among the prosumers thereby encouraging adoption and investments in DERs \cite{wang2016incentivizing}. At the same time, local energy sharing through P2P energy trading reduces peak demand on the main grid thereby reducing investments and operational costs \cite{abdella2018peer}.

Although P2P energy trading has many advantages and is the promising next generation energy management technique for smart grids, the following challenges must be addressed. Firstly, it is generally intractable to develop a P2P energy trading model that is accurate and efficient enough for optimal energy trading decision making \cite{wang2020peer}. Secondly,  it is hard to implement P2P energy trading at a large-scale and in real time  when conventional and model-based optimization techniques are used \cite{lu2019reinforcement, chen2018local, bi2020real}. Thirdly, P2P energy trading has so many uncertainties caused  by the stochastic nature of renewable generation, power consumption and electricity price \cite{gao2020distributed, kim2015dynamic}. Fourthly, as the distribution network acts as a medium for energy exchange during energy trading, its own hard technical constraints including voltage limits and power balance constraints must be satisfied \cite{tushar2020peer}. Finally, as prosumers do not have access to information about others, its difficult to make optimal energy trading decisions. Many of the existing methods e.g. in \cite{guerrero2018decentralized, morstyn2019integrating, kim2019p2p, morstyn2018multiclass, paudel2020decentralized}  are model-based approaches which require domain expert knowledge to  model P2P energy trading, making them difficult to apply.

To address the above challenges, many studies are focusing on the use of deep reinforcement learning (DRL), which is an artificial intelligence framework with proven success in playing Atari and Go games \cite{mnih2015human}. DRL is a combination of deep learning and reinforcement learning \cite{mnih2015human, sutton2018reinforcement}. Compared to model-based methods, DRL-based methods have the following advantages: (i) they are model-free and the agents learn optimal energy trading policies by interacting with the energy trading environment \cite{chen2019realistic, kim2020automatic}. Thus, they can operate without explicit knowledge and rigorous mathematical models  of the environment, (ii) they have self-adaptability and can operate in an on-line way without requiring forecast information about the energy trading environment \cite{cao2020deep, yu2020deep}, and (iii) they are data-driven and capable of determining optimal control actions in real-time even in complex energy trading environments \cite{gao2020distributed}. 

In \cite{chen2019realistic,chen2018local}, Chen \textit{et al.} proposed a DRL-based algorithm  to maximize trading profits while also minimizing the dependence on the power plant. In \cite{kim2020automatic} Kim  and Lee proposed a DRL-based automatic trading algorithm originally designed for stock trading to maximize profits of prosumers participating in P2P energy trading. In \cite{lu2019reinforcement} Lu \textit{et al.} presented a DRL-based  microgrid energy trading scheme to determine  optimal energy trading policy according to  predicted  energy generation, power consumption and battery energy level. In \cite{gao2020distributed}, Gao \textit{et al.} proposed a multi-agent DRL-based approach for minimizing energy costs for P2P energy trading prosumers in a microgrid. Although some DRL-based methods have been proposed in above-mentioned studies, none of them considers the distribution network constraints and loss. As the distribution network acts a medium of exchange for the traded energy, failing to consider its underlying electrical network constraints may cause the outcome of the studied DRL-based energy trading schemes to be impractical.

In this paper, we propose a  multi-agent DRL algorithm for determining optimal energy trading policies that minimize energy costs while satisfying distribution network constraints. Each prosumer is modelled as an agent. The energy trading environment is modelled as a multi-agent environment where an action of one agent affects the actions of others, making the entire energy trading environment to be non-stationary from an agent's perspective. As most  DRL-based methods such as deep Q-networks \cite{mnih2015human} perform poorly in multi-agent settings because they do not use information of other agents during training, we  adopt a multi-agent deep deterministic gradient policy (MADDPG) \cite{lowe2017multi} based framework to design the proposed algorithm. With the proposed MADDPG-based algorithm, training is centralized with each agent using states and actions of other agents. This makes the environment to be stationary during training even as the agent actions change. Meanwhile execution is decentralized with each agent using only local information to make actions without knowing others' information. The main contributions of this paper are summarized as follows:
\begin{itemize}
	\item  Propose a  local P2P energy trading market which enables a distribution system operator (DSO) to leverage prosumers' battery energy storage system as a flexible asset to satisfy the distribution network constraints. 
	\item Propose a MADDPG-based algorithm  to learn optimal energy trading policies for each prosumer to minimize the energy costs  while satisfying the distribution network constraints.
	\item Design actor and critic networks for each agent to ensure that training of the agents is stable and that the output from the actor network is optimal.
	\item Introduce a novel strategy using distribution network tariffs (DNT) to incentivize the prosumers to provide the  flexibility required to satisfy the network constraints.  
\end{itemize}
	
The rest of the paper is organised as follows. Section \ref{sec:system model} presents the proposed P2P energy trading model and the energy cost minimization problem considered in this paper. Section \ref{sec:MADRL} presents the proposed algorithm. Simulation results that verify the effectiveness of the proposed algorithm are given in Section \ref{case study}. Section \ref{conc} concludes the paper. 

\section{System Model}\label{sec:system model}
\begin{figure}[!t]
	\centering
	\includegraphics[width=2.5in]{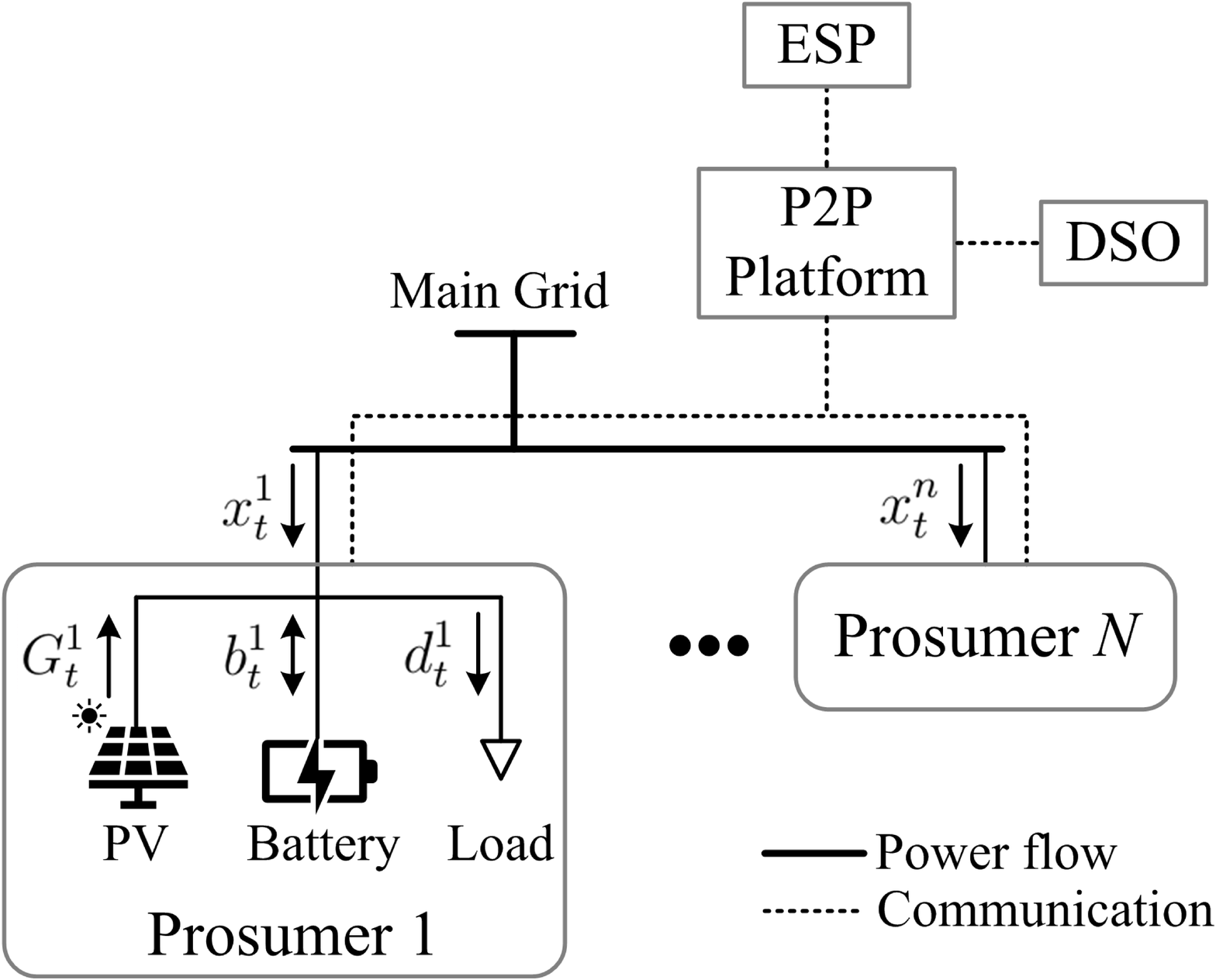}
	\caption{Distribution network model with P2P platform.}
	\label{SN}
\end{figure}

Fig. \ref{SN} shows a simplified distribution network and P2P energy trading model studied in this paper. The distribution network can be described as a connected graph, \( \mathcal{G} = \left(\mathcal{N}, \mathcal{E}\right)\) where \( \mathcal{N} = \{0,1,\dots,N\}\) is a set of buses and \( \mathcal{E}\) is a set of distribution lines. Thus, each bus, \(n\in \mathcal{N}\) corresponds to a prosumer (e.g. a household, business or commercial building) connected to the distribution network. Bus \(n = 0\) represents a substation bus, which links the distribution network to the main grid, and no consumer or prosumer is connected to it. For easy of reference, we use the term prosumer to refer to either a consumer or prosumer in the remaining part of the paper, unless explicitly stated. It should be noted that not all prosumers connected to the distribution network are willing to participate in the P2P energy trading. We denote the set of prosumers participating in P2P energy trading as \(\mathcal{P} = \{1,\dots, P\}\). Each prosumer, \(p\in \mathcal{P}\) consists of a solar PV system, battery and/or load.  In addition, each prosumer is equipped with an energy management system (EMS) to: (i) collect and send to the P2P platform data such as PV generation, energy consumption, battery charge and discharge power, (ii) receive the price signal from the P2P platform and (iii) optimally schedule the battery energy storage system.

As energy buying and selling prices via the P2P platform must be lower and higher than the import and export tariffs set by the ESP respectively \cite{long2018peer}, the P2P platform communicates with the ESP in achieving this. To avoid overloading the distribution network, the energy sharing between the prosumers must satisfy distribution network constraints such as voltage limits and losses. In many countries including the UK, management and operation of the distribution network is a responsibility of the DSO, who operates independently from the P2P platform \cite{tushar2020peer,olivella2018optimization}. To satisfy the network constraints at every transaction, the intended power exchange between the prosumers must be approved by the DSO either through penalties or monetary incentives \cite{morstyn2019integrating, kim2019p2p}. Bi-directional communication links are required for communication between the prosumer EMS, P2P platform, DSO and the ESP. Further, we assume that the prosumer EMS, P2P platform, ESP and DSO operate on a common time horizon, \(\mathcal{T} = \{1,\dots, T\}\) with equal time slots, \(\Delta t\).

\subsection{Prosumer Model}\label{subsec:prosumer model}
The PV generation profile for prosumer \(p\) during the operation horizon \(\mathcal{T}\) is defined as follows
\begin{equation}
	\mathbf{G}^p = \{G_1^p, G_2^p,\dots, G_T^p\}\,,\;\;\;p \in \mathcal{P}
\end{equation}
We assume that the PV is operated in maximum power point tracking (MPPT) mode \cite{esram2007comparison} and thus, \(\mathbf{G}^p\)  is maximum power output. 

The total power consumption profile of  prosumer \(p\) during time horizon \(\mathcal{T}\) can be defined as follows
\begin{equation}
\mathbf{D}^p = \{d_1^p, d_2^p,\dots, d_T^p\}\,,\;\;\;p \in \mathcal{P}
\end{equation}
We consider loads that do not have a certain amount of flexibility. Consideration of flexible loads is beyond the scope of this paper and considered as future work.

For prosumer \(p\in \mathcal{P}\), let \(SoC_{t}^p\)  be the battery state of charge (SoC), which indicates the amount of energy remaining in the battery after a charge or discharge operation. Let  \(b_t^p\) be the battery power output (positive \(b_t^p\) to denote discharging and negative \(b_t^p\) to denote charging), \(\eta_t^p\) be the battery charge or discharge efficiency and let \(E_b^p\) be the battery energy capacity.  The dynamics of the SoC can be modelled as follows \cite{cao2020deep}
\begin{equation}
	SoC_{t + \Delta t}^p = SoC_{t}^p - \frac{\eta_t^pb_t^p\Delta t }{E_b^p}\,,\;\;p\in \mathcal{P}\,,\;\;t\in \mathcal{T}
	\label{SoC}
\end{equation}
It should be noted that the value of \(\eta_t^p\) is calculated differently based on whether the battery is charging or discharging \cite{cao2020deep}.
To prolong the battery lifetime, \(b_t^p\) must be restricted  within a certain range as follows
 
\begin{equation}
\frac{E_b^p\left(SoC_{t}^p -SoC_{max}^p\right)}{\eta_t^p\Delta t} \le b_{t}^p \le \frac{E_b^p\left(SoC_{t}^p - SoC_{min}^p\right)}{\eta_t^p\Delta t}
\label{SoC_limits}
\end{equation}
\begin{equation}
p\in \mathcal{P}\,,\;\;t\in \mathcal{T} 
\end{equation}
where \(SoC_{min}^p\) and \(SoC_{max}^p\) are predetermined SoC limits to indicate a fully discharged and charged battery respectively.

Also, \(b_t^p\) must satisfy the  power limits of the inverter to which the battery is connected as follows
\begin{equation}
	b_{min}^p \le b_t^p \le b_{max}^p\,,\;\;p\in \mathcal{P}\,,\;\;t\in \mathcal{T}
	\label{b_limits}
\end{equation}
where \(b_{min}^p\) and \(b_{max}^p\) are  minimum  and maximum inverter power limits respectively.

Practically, the lifetime of the battery is shorter than any other asset in the distribution network. Thus, its wear cost has great impact on the economics of the energy trading strategies of the prosumers.  The empirical wear cost \(\varpi^p\)  of the battery can be expressed as \cite{han2014practical}
\begin{equation}
\varpi^p = \frac{C_b^p}{ACC \times 2\times DoD \times E_b^p \times \mu_b^2}
\end{equation}
where $C_b^p$ is battery price per kWh, $DoD$ is depth of discharge at which the battery is cycled, $\mu_b$ is round trip efficiency and $ACC$ is life cycle at a specific $DoD$. $ACC$ is multiplied by two as one cycle consists of charge and discharge phases.

As profiles of power generation and consumption are different from each other,  prosumer \(p\) can assume the role of an energy buyer or seller  at any time \(t\in \mathcal{T}\) based on the net power \(x_t^p\) which is defined as follows
\begin{equation}
	x_t^p = d_t^p - \left(G_t^p + b_t^p\right)\,,\;\;p\in \mathcal{P}\,,\;\;t\in \mathcal{T}
	\label{net_power}
\end{equation}
That is, if \(x_t^p \ge 0\), the prosumer is an energy buyer, buying energy from others or the grid to meet its power deficit. The prosumer is an energy seller if \(x_t^p < 0\), selling the excess energy to others or the grid. 

\subsection{P2P Pricing Mechanism}\label{subsec:price_mechanism}
Energy buying and selling all happens through the P2P platform as shown in Fig. \ref{SN}.  To set the energy buying/selling price in the platform, we adopt the suppy-to-demand ratio (SDR) based pricing mechanism \cite{liu2017energy,long2018peer,wang2020peer}, mainly for two reasons: (i) it is simple in principle and easy to obtain, and can be updated in real time
and (ii) it satisfies the basics of modern economics, i.e., price is inversely proportional to SDR as demonstrated in the following paragraphs. 

SDR is defined as the ratio of the total power supply to the total demand in the energy sharing community, i.e.,
\begin{equation}
SDR^t = \frac{\sum_{p\in \mathcal{P}} \left(G_t^p +  b_t^p\right)}{\sum_{p\in \mathcal{P}} d_t^p}\,,\;\;t\in \mathcal{T}
\label{SDR}
\end{equation}
The SDR varies with time because of the volatility of solar generation and power consumption. This means that the energy buying/selling price is also not constant but fluctuating according to the SDR. Let the ESP's import and export prices as received by the P2P platform  be \(\lambda_{b}^t\) and \(\lambda_s^t\) respectively, where \(\lambda_{b}^t \ge \lambda_s^t\). Let the prosumer's buying and selling prices through the P2P platform be $\pi_{b}^t$ and $\pi_{s}^t$ respectively, where $\pi_{b}^t \le \lambda_{b}^t$ and $\pi_{s}^t \ge \lambda_{s}^t$. The buying/selling price vector set by the P2P platform can be defined as follows
\begin{equation}
	\mathbf{\pi} = \{\pi_{b}^1, \pi_{b}^2,\dots, \pi_{b}^T\;:\;\pi_{s}^1, \pi_{s}^2,\dots, \pi_{s}^T\}
	\label{p2p_price}
\end{equation}
The $\pi_{b}^t$ and $\pi_{s}^t$ can be obtained as a function of \(SDR^t\), \(\lambda_{b}^t\) and \(\lambda_s^t\) as follows \cite{long2018peer}
\begin{equation}
\pi_{s}^t = 
\begin{cases}
\frac{\left(\lambda_{s}^t + \lambda\right)\lambda_{b}^t}{\left(\lambda_{b}^t - \lambda_{s}^t -\lambda\right)SDR^t + \lambda_{s}^t + \lambda},\;\;0\le SDR^t \le 1\\
\lambda_{s}^t + \frac{\lambda}{SDR^t} \;\;\;SDR^t > 1\\
\end{cases}
\label{sell_price}
\end{equation}
\begin{equation}
\pi_{b}^t = 
\begin{cases}
\pi_{s}^tSDR^t + \lambda_{b}^t\left(1-SDR^t\right),\;\;0\le SDR^t\le 1\\
\lambda_{s}^t + \lambda\;\;\;\;SDR^t > 1
\end{cases}
\label{buy_price}
\end{equation}
where $\{\lambda|0\le \lambda \le \left(\lambda_{b}^t - \lambda_{s}^t\right)\}$ is a compensation price which is used to incentivize prosumers to continue participating in P2P energy trading when $SDR^t > 1$ (the situation which happens when prosumers have more power supply than demand). Without the compensation price, buying price would be equal to selling price when $SDR^t > 1$, a situation that would favour prosumers who are buyers and not sellers. This may discourage the sellers from participating in P2P energy trading especially during periods of high PV generation.  

The $SDR^t$ given by (\ref{SDR}) mainly depends on the adjusted power consumption and battery charge and discharge power from all the prosumers. This means that \(\pi_{s}^t\) and \(\pi_{b}^t\) largely depend on the choice of \(b_t^p\). Decreasing \(b_t^p\) (when charging the battery) drives $SDR^t$ towards zero and \(\pi_{s}^t\) or \(\pi_{b}^t\)  towards \(\lambda_{b}^t\). Conversely, increasing \(b_t^p\) (when discharging the battery) drives $SDR^t$ towards 1 and \(\pi_{s}^t\) or \(\pi_{b}^t\)  towards \(\lambda_{s}^t\). Thus, price is inversely proportional to the SDR. In both cases the following relationship is satisfied
\begin{equation}
	\begin{cases}
	\pi_{b}^t \le \lambda_{b}^t\\
	\pi_{s}^t \ge \lambda_{s}^t	
	\end{cases}
\end{equation}
That is, prosumers are better off buying and selling their energy via the P2P platform because of lower buying prices and higher selling prices compared to  the prices, \(\lambda_{b}^t\) and \(\lambda_{s}^t\) offered by the ESP. 

\subsection{Distribution Network Tariffs}\label{sec:trading method}
The selling and buying price given by (\ref{sell_price}) and (\ref{buy_price}) respectively do not take distribution network constraints such as line congestion and voltage limits into account. If not controlled, the net power obtained for each energy trading transaction  may overload the distribution network. We introduce the use of a distribution network tariff (DNT) to incentivize transactions that do not violate the distribution network constraints and penalize those that violate the constraints. 

Prosumer's contribution towards violation of network constraints  depends on its location on the distribution network and operational time \cite{papavasiliou2017analysis}. Thus, we derive the DNTs from distribution locational marginal pricing (DLMP), a temporal-spatio pricing mechanism which exposes prosumers to the true cost of energy delivery in the distribution network \cite{papavasiliou2017analysis, bai2017distribution}. The DLMP can be decomposed into four constituent components; marginal cost of energy demand, marginal cost of network loss, marginal cost of congestion and marginal cost of bus voltage \cite{papavasiliou2017analysis, kim2019p2p}.  As the energy buying/selling price  (\ref{p2p_price}) is set by the P2P platform, the proposed DNTs are determined from the other three components of the DLMP, i.e. the  marginal cost of network losses, marginal cost of congestion and  marginal cost of voltage. Calculation of the DLMP is detailed in \cite{papavasiliou2017analysis}. 

Let the DLMP obtained at time \(t,\;\;t \in \mathcal{T}\) for the substation bus (i.e., $n=0,\;\;n \in \mathcal{N}$)  be \(\lambda_0^t\) and the DLMP for prosumer \(p\) be \(\lambda_p^t\). The DNT \(\delta_p^t\) for prosumer \(p\) can be calculated as follows
\begin{equation}
	\delta_p^t = \lambda_p^t - \lambda_0^t\,,\;\;\;p \in \mathcal{P}\,,\;\;t \in \mathcal{T}
	\label{DNT}
\end{equation}
As \(\lambda_0^t\) only accounts for the marginal cost of energy delivery at the substation bus (which is the same at every bus \(n \in \mathcal{N}\)), subtracting it from \(\lambda_p^t\) gives \(\delta_p^t\), which is the sum of marginal cost of line losses, congestion and voltage. \(\delta_p^t\) is zero when the net power injected by prosumer \(p\) at \(t \in \mathcal{T}\)  does not cause network losses, congestion and/or violate voltage limits. Otherwise, \(\delta_p^t\) is not equal zero due to either network loss, congestion and/or voltage limit violation.

As \(\delta_p^t\)  reflects the condition of the entire distribution network considering both location and time, it is therefore a suitable
tariff to manage the distribution network constraints. \(\delta_p^t\) can be used as an incentive to incentivize a prosumer whose net power transfer helps to satisfy the network constraints and penalize the one whose power transfer violates the constraints. 

\subsection{Problem Formulation}\label{prob_formulation}

Each prosumer \(p\) that can schedule the operation of its energy assets can be considered to be an agent. Thus, P2P energy trading can be described as a multi-agent system. Each agent's energy trading decision at a given time slot, $t$  depends on the current information it receives from its assets (e.g. battery energy level) and the P2P platform (e.g. energy buying/selling price and DNT), and not on the prior history. Thus, energy trading and the subsequent scheduling of the energy assets can be formulated as a Markov decision process (MDP) \cite{littman1994markov} with continuous action spaces (i.e. assuming that operation of the assets e.g., battery is continuous). 

Let the set of agents be the same as that of prosumers (i.e. \(p \in \mathcal{P}\)). The MDP for each agent \(p\) proceeds as follows: Given a local agent state \(s_p^t \in \mathcal{S}\)  at time slot $t$, where \(s_p^t = (G_t^p, d_t^p, SoC_t^p)\), the agent selects an action \(a_p^t \in \mathcal{A}\), where  \(a_p^t = (b_t^p)\) based on a stochastic policy, \(\pi_{\theta_p}\). The taken action takes the agent into a next local state \(s_p^{t'} \in \mathcal{S}\) according to a state transition probability function, \(\mathcal{F}\). At the end of the time slot,  the agent receives a reward, \(r_p^t\) as a function of the current state and action as follows
	
\begin{equation}
r_p^t = -\sum_{t\in \mathcal{T}}\left[\left(\pi + \delta_p^t\right)x_t^p + \varpi^p|b_t^p|\right]\Delta t
\label{reward}
\end{equation} 
\begin{equation*}
\pi = 
\begin{cases}
\pi_{b}^t\,,\;\;\;\text{if}\;\;x_t^p \ge 0\\
\pi_{s}^t\,,\;\;\;\text{Otherwise}
\end{cases}
\end{equation*}
where the first term is the cost for both purchasing energy in the P2P platform and using the distribution network. The second term is the cost of using the battery. 

It is important to note that \(\delta_p^t\) is equal to zero when the net power \(x_t^p\) does not contribute to network losses, congestion and/or voltage limit violations, otherwise \(\delta_p^t\) is non-zero. Furthermore, through  \(\pi\) and \(\delta_p^t\), the reward is a function of all agent states and actions in the network (i.e. to determine \(\pi\) and \(\delta_p^t\), the P2P platform and the DSO needs access to all the states and actions of all agents in the network). In other words, an action of one agent affects the rewards of all other agents in the system.
	
The goal of each agent is to maximize its own expected reward \(R_p = \sum_{t=0}^{T}\gamma^tr_p^t\) where \(\gamma\) is a discount factor. As market prices, generation and demand are volatile in nature, it is generally impossible to obtain with certainty the state transition probability function \(\mathcal{F}\)  required to derive an optimal policy \(\pi_{\theta_p}\) needed to maximize \(R_p\). To circumvent this difficulty, we propose to use an artificial intelligence-based approach which is data-driven and model-free as discussed in Section \ref{sec:MADRL}.
		
\section{Proposed  Learning Algorithm}\label{sec:MADRL}

\begin{figure*}[!t]
	\centering
	\includegraphics[width=\textwidth]{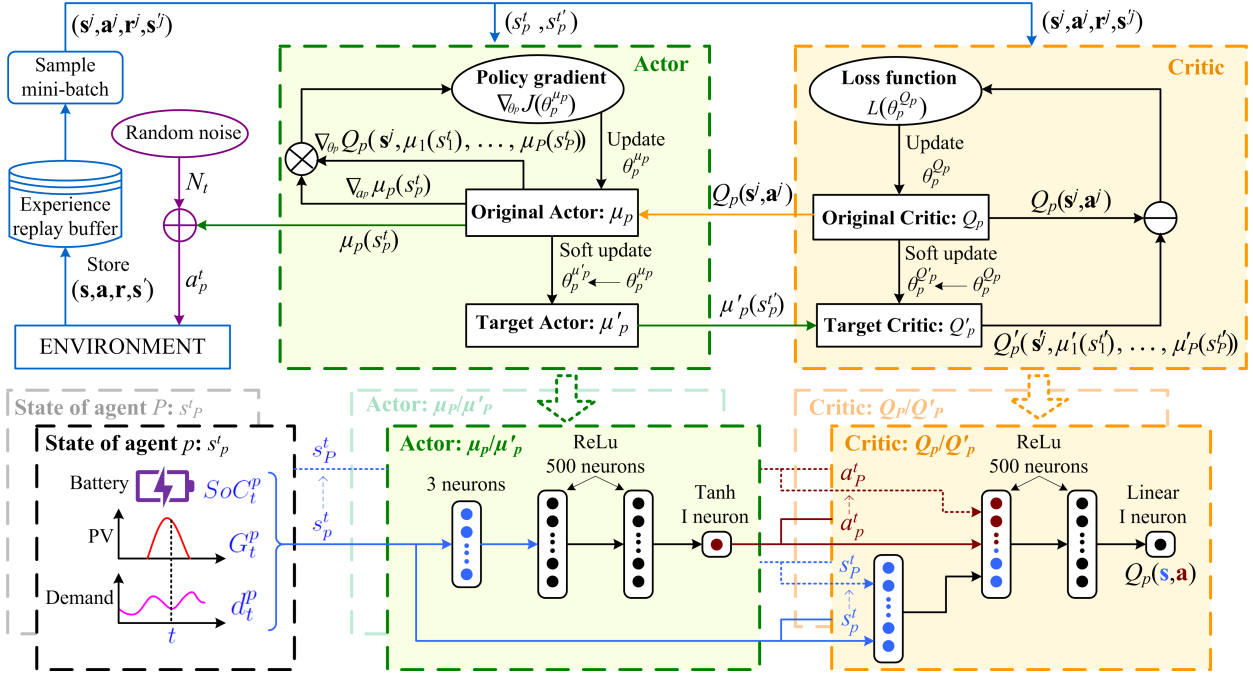}
	\caption{Architecture and workflow  of the proposed MADDPG algorithm. Each agent \(p\), \(p = 1,2,\dots, P\) consists of an original actor network \(\mu_p\) (and target actor network \(\mu_p^{'}\)) and original critic network \(Q_p\)  (and target critic network \(Q_p^{'}\) ).}
	\label{MADDPG_drawing}
\end{figure*}
\subsection{Deep Reinforcement Learning}\label{sec:DRL}
Reinforcement learning (RL) is the process in which agents learn for themselves through trial and error \cite{sutton2018reinforcement} the optimal policy \(\pi_{\theta_p}\) to achieve optimal actions that maximize the cumulative reward \(R_p\). Like a human, agents need to construct and learn their own knowledge directly from raw data such as a historic solar PV generation, demand and market prices. This can be achieved by DRL. DRL has given rise to several value-based algorithms such as Deep Q-networks (DQN) \cite{cao2020deep, mnih2015human} and policy-based algorithms such as deep deterministic policy gradient (DDPG) \cite{silver2014deterministic, lillicrap2015continuous}.

As each agent's reward as given by (\ref{reward}) depends on actions from other agents, the interaction between the agents during energy trading can be described as a mixed cooperative-competitive. Naive application of DQN and policy gradient algorithms to such multi-agent settings performs poorly because they do not use information of other agents during training. We propose to use an MADDPG-based algorithm which overcomes this difficulty by using states and actions of other agents during training.

\subsection{Multi-Agent Deep Deterministic Policy Gradient Algorithm }\label{sec:MADDPG}

Fig. \ref{MADDPG_drawing} shows the architecture and workflow of the proposed MADDPG algorithm. Each agent is modelled as a DDPG agent, where, however, states and actions are shared between the agents during training. In particular, each agent consists of two networks: an actor network and a critic network. Both actor and the critic networks are created from dense layers with hidden layers having ReLu activations. An actor-network maps the local state of an agent  to optimal actions using a Tanh activation function in the output layer. A critic network evaluates the actions  received from the actor network to improve the performance of the actor network. The output layer of the critic network is activated by a linear function. 

During training, the actor network uses only the local state to calculate the  actions while the critic network uses states and actions of all agents in the system in evaluating the local action. As actions of all agents are known by each agent's critic network, the entire environment is stationary during training. During execution, critic networks are removed and only actor networks are used. This means that with MADDPG, training is centralized while execution is decentralized. As DNTs are obtained independently by the DSO, we leverage the centralized training provision of the MADDPG to incorporate the DNTs in the algorithm during training.

\subsection{Proposed Algorithm }\label{sec:algorithm}

The details of the proposed MADDPG algorithm which are illustrated in Fig. \ref{MADDPG_drawing} are given by Algorithm \ref{algorithm}. Let the actor and critic network of agent \(p\) be denoted as \(\mu_p\) and \(Q_p\), and the associated network weights as \(\theta_p^{\mu_p}\) and \(\theta_p^{Q_p}\) respectively. 
Before training starts, \(\mu_p\) and \(Q_p\) (which we refer to as original networks) are created and their weights  $\theta_p^{\mu_p}$ and $\theta_p^{Q_p}$ are randomly initialized.  To add stability to the training,  target actor $\mu_p^{'}$ and target critic $Q_p^{'}$ networks which are identical to the original networks \(\mu_p\) and \(Q_p\) are also created and their weights are initialized as $\theta_p^{\mu_p'} \leftarrow \theta_p^{\mu_p}$ and $\theta_p^{Q_p'} \leftarrow \theta_p^{Q_p}$. 

For each agent \(p\), a replay buffer \(\mathcal{D}\) is  created and initialized  to store list of tuples \((\mathbf{s}, \mathbf{a}, \mathbf{r}, \mathbf{s}^{'})\) known as experiences, where \(\mathbf{s} = \left(s_1^t,\dots, s_P^t\right)\), \(\mathbf{a} = \left(a_1^t,\dots, a_P^t\right)\), \(\mathbf{r} = \left(r_1^t,\dots, r_P^t\right)\) and $\mathbf{s}^{'} = (s_1^{t'},\dots, s_P^{t'})$. The replay buffer adds stability to the training as agents learn by sampling mini-batches from all of the accumulated experiences during training.

For each  training episode,  a random process for action exploration and an  initial state \(\mathbf{s}\) are initialized. We use Ornstein-Uhlenbeck process \cite{uhlenbeck1930theory} for generating the noise \(\mathcal{N}_t\) for action exploration. With the received state \(s_p^t\),  \(s_p^t \in \mathbf{s}\), and noise \(\mathcal{N}_t\), each agent makes an action given by
\begin{equation}
	a_p^t = \mu_p\left(s_p^t\right) + \mathcal{N}_t
	\label{action}
\end{equation}
where \(\mu_p\left(s_p^t\right)\) is  output (action) of the actor network \(\mu_p\). 

The actions from the agents together with their states at time slot, \(t\) are used to simulate  the energy trading mechanism including the calculation of selling/buying price and the DNTs. At the end of the time slot, each agent calculates its own reward \(r_p^t\), $r_p^t \in \mathbf{r}$ given by (\ref{reward}) and observes a new state $s_p^{t'}$,  $s_p^{t'} \in \mathbf{s}^{'}$. The experience $(\mathbf{s}, \mathbf{a}, \mathbf{r}, \mathbf{s}^{'})$ is stored in the replay buffer \(\mathcal{D}\) and the initial state $\mathbf{s}$ gets updated; $\mathbf{s} \gets \mathbf{s}^{'}$.

For each agent \(p\), the actor \(\mu_p\) and critic \(Q_p\) networks are trained by (random) sampling \(S\) number of transitions from the replay buffer \(\mathcal{D}\). The transitions are used to update the network weights for both original and target actor and critic networks. Let $(\mathbf{s}^j, \mathbf{a}^j, \mathbf{r}^j, \mathbf{s}^{'j})$ be an experience for each transition \(j\), \(j \in S\). Each agent \(p\) updates the  weights of its original critic network (i.e. $\theta_p^{Q_p}$)  by minimizing the loss 
\begin{equation}
L\left(\theta_p^{Q_p}\right) = \frac{1}{S}\sum_{j=1}^{S}\left(y_p^j - Q_p\left(\mathbf{s}^j, \mathbf{a}^j\right)\right)^2
\label{critic_loss}
\end{equation} 
where $Q_p\left(\mathbf{s}^j, \mathbf{a}^j\right)$ is the predicted output of the original critic network and \(y_p^j\) is its target value which is given by
\begin{equation}
y_p^j = r_p^{t^j} + \gamma Q_p^{'} \left(\mathbf{s}^{'j}, a_1^{t'}, \dots, a_P^{t'}\right)\Big|_{a_p^{t'} = \mu_p^{'} \left(s_p^{t'}\right),\;\; p \in \mathcal{P}}
\end{equation}
where \(a_p^{t'} = \mu_p^{'} \left(s_p^{t'}\right)\) is the predicted action  by the target actor network and  \(Q_p^{'} \left(\mathbf{s}^{'j}, a_1^{t'}, \dots, a_P^{t'}\right)\) is the predicted value by the target critic network.

Weights for the original actor network (i.e. $\theta_p^{\mu_p}$) are updated using sampled policy gradient
\begin{equation}
	\nabla_{\theta_p^{\mu_p}}J\left(\theta_p^{\mu_p}\right) = 
	\nabla_{\theta_p^{\mu_p}}\mu_p\left(s_p^t\right)
	\nabla_{a_p^t}Q_p\left(\mathbf{s}^j,a\right)
	\label{policy_grad}	
\end{equation}
where \(a = \left(\mu_1\left(s_1^t\right),\dots,\mu_P\left(s_P^t\right) \right)\).

Weights for both target actor and critic network (i.e. $\theta_p^{\mu_p'}$ and $\theta_p^{Q_p'}$) are updated as follows
\begin{equation}
	\begin{cases}
	\theta_p^{Q_p'} \gets \tau \theta_p^{Q_p} + \left(1 - \tau\right)\theta_p^{Q_p'}\\
	\theta_p^{\mu_p'} \gets \tau \theta_p^{\mu_p} + \left(1 - \tau\right)\theta_p^{\mu_p'}	
	\end{cases}
	\label{target_update}
\end{equation}
where \(\tau\) is the learning rate.

After training, the trained critic network and the replay buffer are removed from each agent. Let the trained actor network for  agent \(p\) be \(\mu_p^{*}\). At every time slot $t$, each agent \(p\) only requires to make an observation of the local state \(s_p^t\) to obtain optimal actions \(a_p^t = \mu_p^{*} \left(s_p^{t}\right)\). The obtained actions are considered to be optimal for energy trading and for satisfying the distribution network constraints. 
\begin{algorithm}
	\caption{MADDPG Algorithm for P2P Energy Trading.}
	\begin{algorithmic}[1]
		\STATE Randomly initialize (original) actor and critic networks 
		\STATE Initialize (target) actor and critic networks 	
		\STATE Initialize replay buffer \(\mathcal{D}\) 	
		\FOR {epidode = 1 to \(M\)}
		\STATE Initialize a random process $\mathcal{N}_t$ for action exploration
		\STATE Observe initial state \(\mathbf{s}\)
		\FOR {$t$ = 1 to \(T\)}
		\STATE For each agent \(p\), make an action according to (\ref{action})
		\STATE Execute the actions \(\mathbf{a}\), calculate the reward \(\mathbf{r}\) using (\ref{reward}) and observe next states \(\mathbf{s}^{'}\)
		\STATE Store \((\mathbf{s}, \mathbf{a}, \mathbf{r}, \mathbf{s}^{'})\) in \(\mathcal{D}\) 
		\STATE Update $\mathbf{s}$ $\gets$  $\mathbf{s}^{'}$
		\FOR {agent \(p = 1\) to \(P\)}
		\STATE Randomly sample $S$ from \(\mathcal{D}\) 
		\STATE Update (original) critic network by minimizing (\ref{critic_loss})
		\STATE Update (original) actor network using policy gradient (\ref{policy_grad})
		\STATE Update (target) actor and critic network by (\ref{target_update})
		\ENDFOR
		\ENDFOR
		\ENDFOR
	\end{algorithmic}
	\label{algorithm}
\end{algorithm}

\section{Case Study}\label{case study}

\begin{figure}[!t]
	\centering
	\includegraphics[width=3.3in]{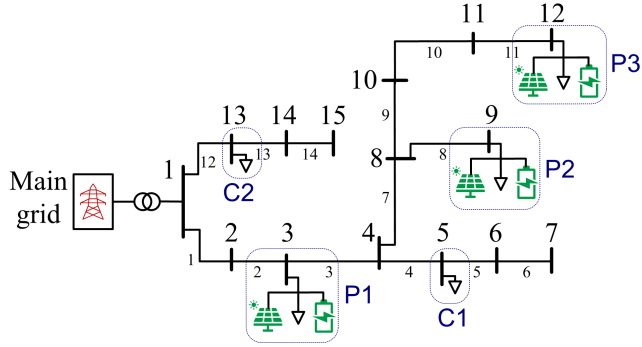}
	\caption{A low voltage 15-bus  radial distribution network with C1 and C2 as consumers and P1, P2 and P3 as prosumers participating in a P2P energy trading scheme.}
	\label{network}
\end{figure}

\subsection{Simulation Parameters}

\begin{figure*}
	\subfloat[Consumer C1]{\includegraphics[width=2.1in]{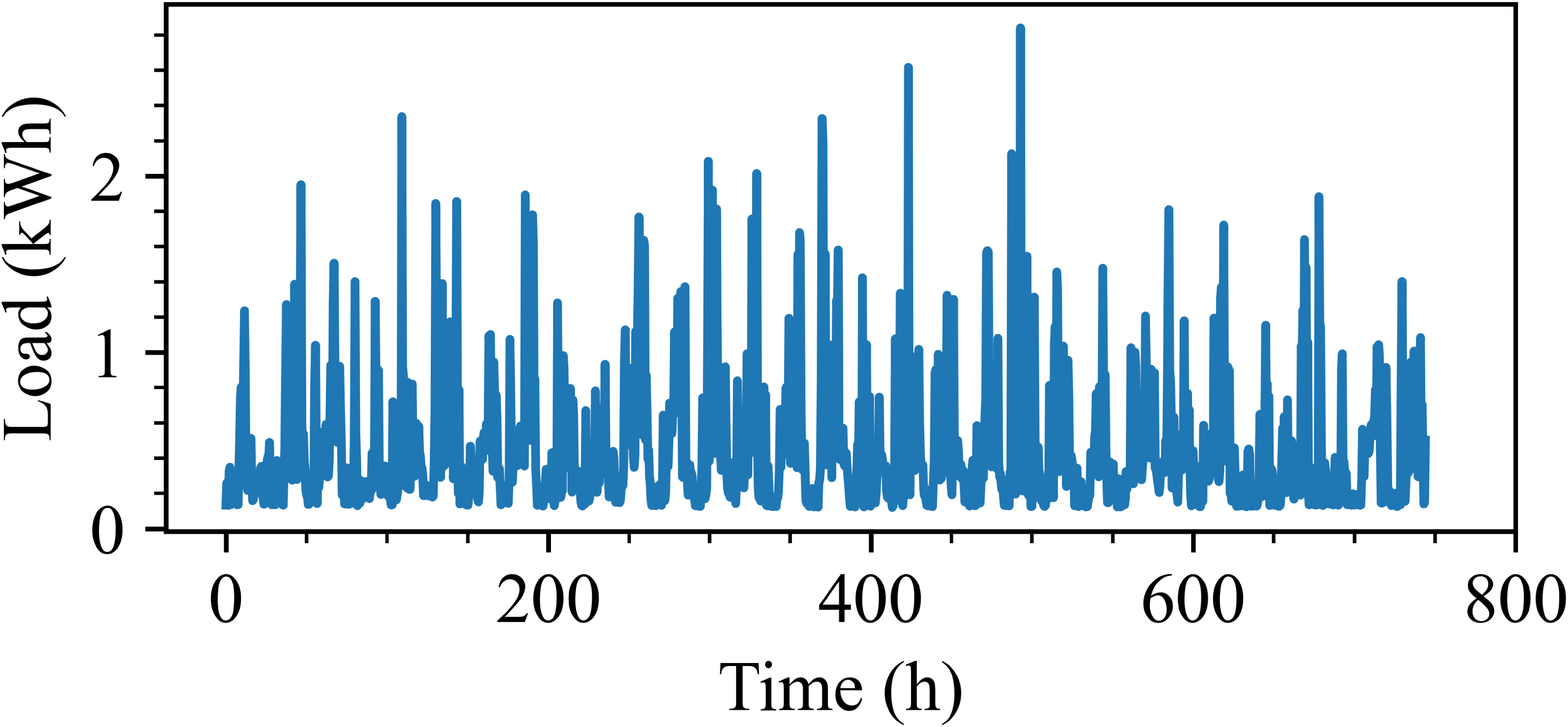}} \hfil
	\subfloat[Consumer C2]{\includegraphics[width=2.1in]{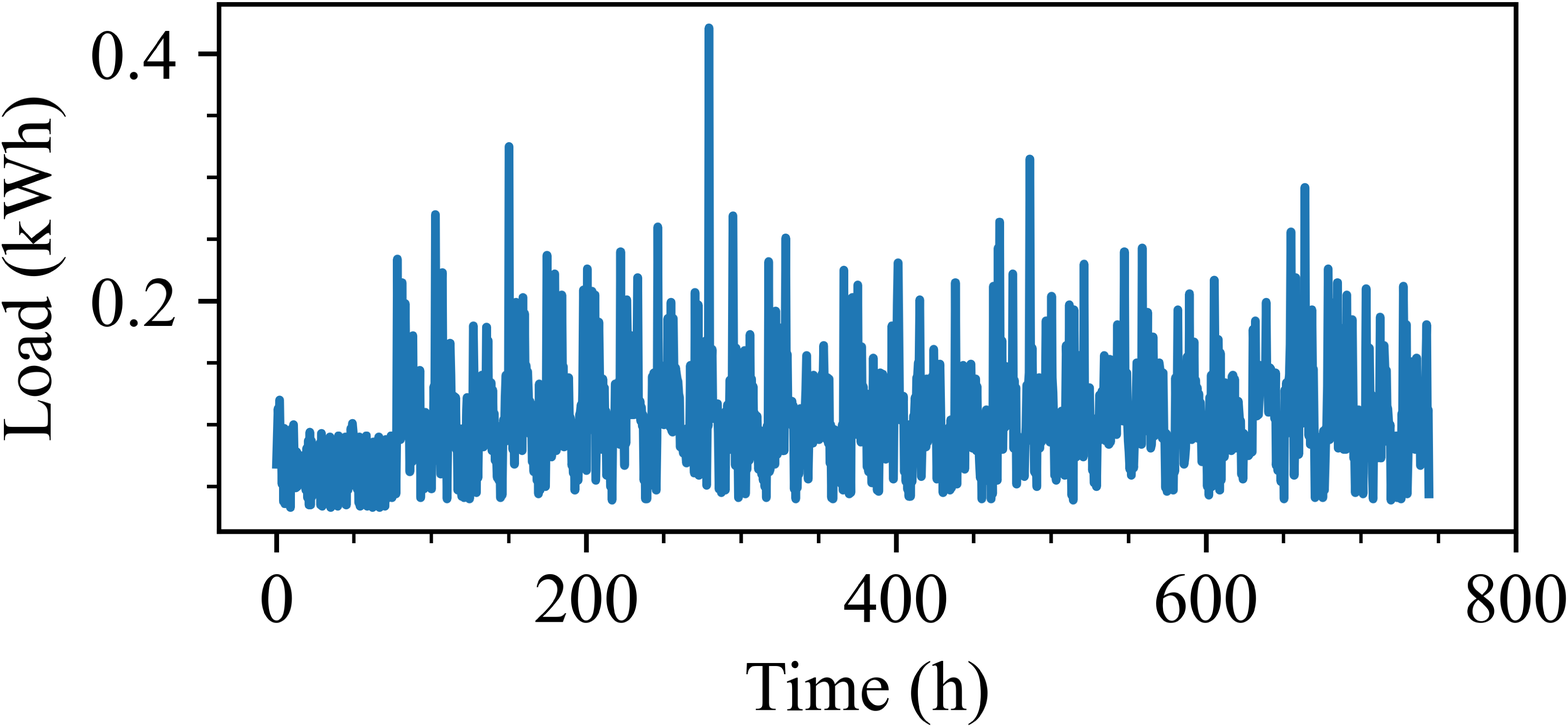}} \hfil 
	\subfloat[Prosumer P1]{\includegraphics[width=2.1in]{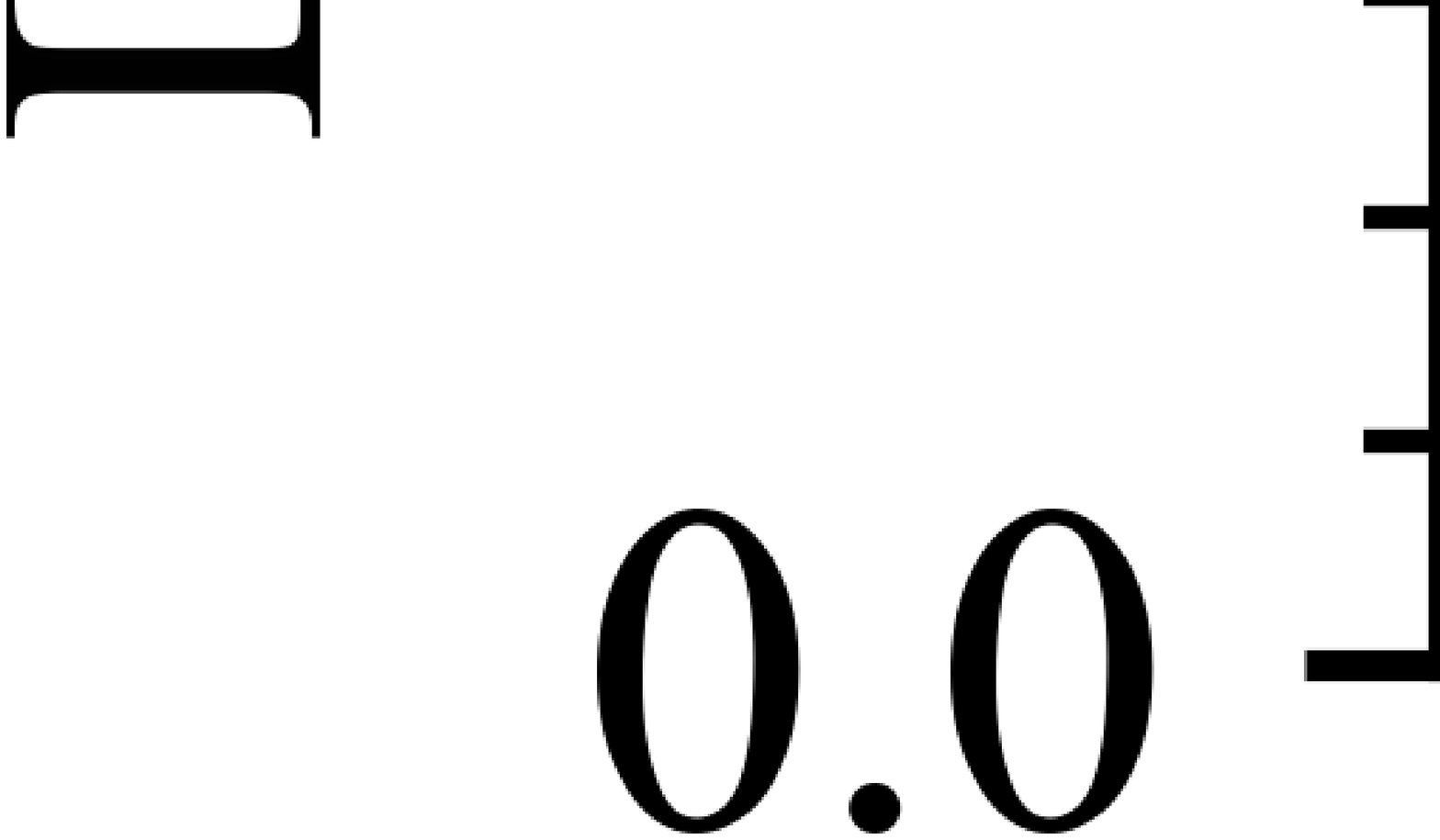}} \hfil 
	\subfloat[Prosumer P2]{\includegraphics[width=2.1in]{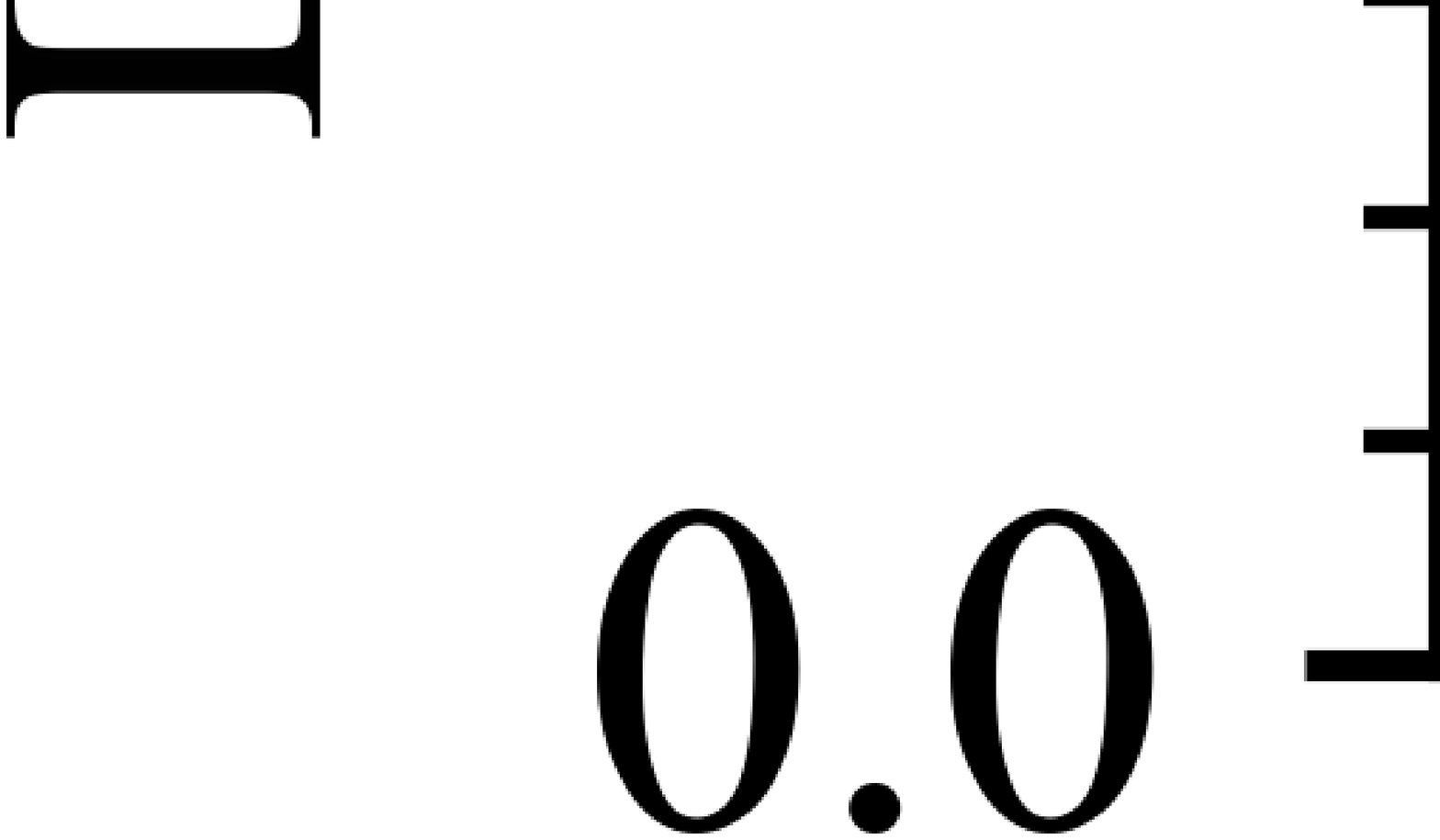}} \hfil
	\subfloat[Prosumer P3]{\includegraphics[width=2.1in]{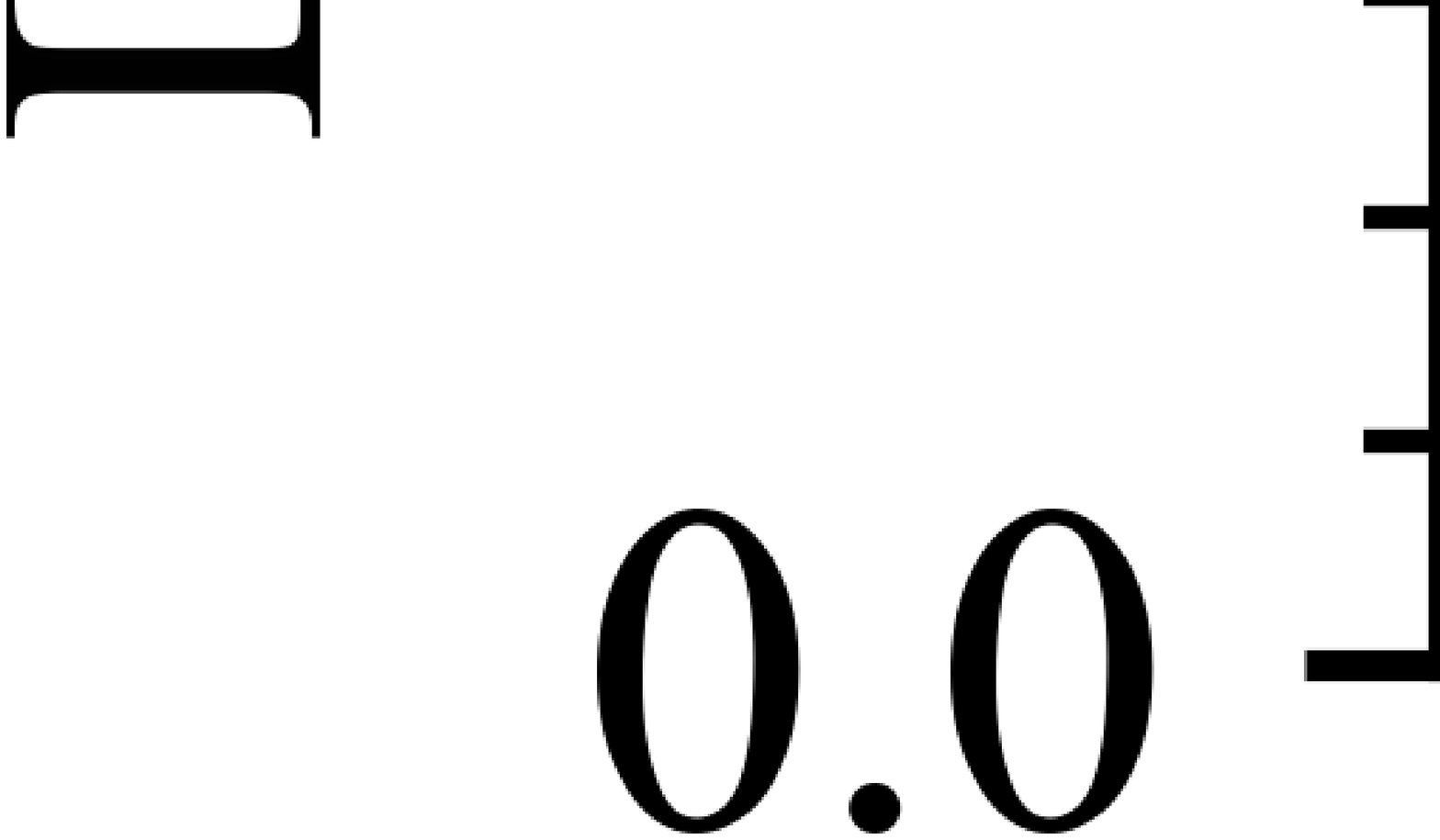}} \hfil
	\subfloat[Prosumer P1, P2 \& P3]{\includegraphics[width=2.1in]{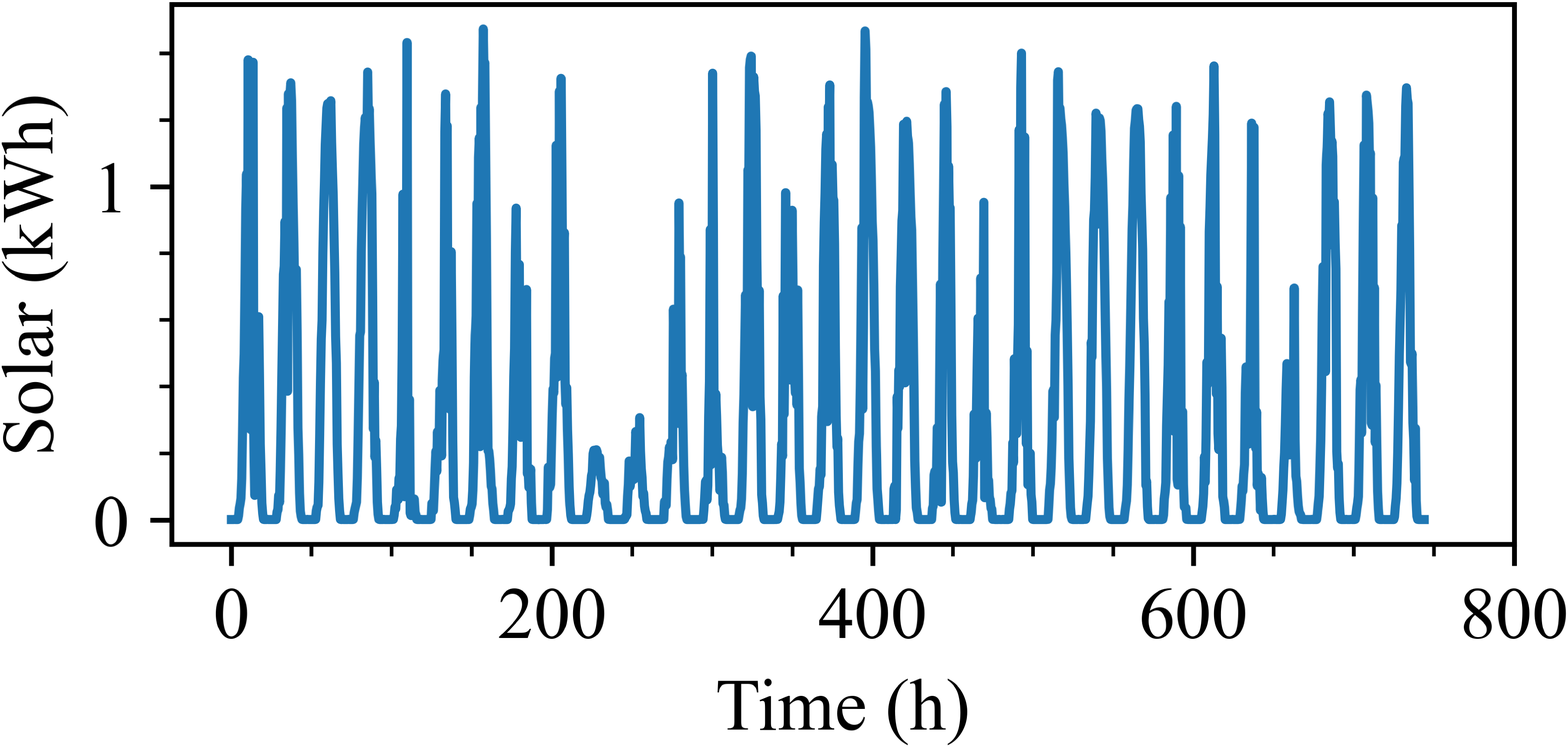}} \hfil 
	\caption{Half-hourly load and solar power profiles for the consumers and prosumers participating in P2P energy trading.}
	\label{profiles}
\end{figure*}

Fig. \ref{network} shows a low voltage 15-bus radial distribution  network \cite{papavasiliou2017analysis} which is used to demonstrate the effectiveness of the proposed algorithm for reducing energy costs while satisfying network constraints. The distribution network parameters are given in \cite{papavasiliou2017analysis}. Two consumers (denoted as C1 and C2) and three prosumers (denoted as P1, P2, and P3)  are considered to participate in the P2P energy trading scheme. These  have  time-varying load and solar power profiles (with 30 minutes resolution) instead of fixed ones  as shown in Fig. \ref{profiles}. The  load and solar power profiles are for one month (744 hours) and  they are obtained from UK's customer led network revolution (CLNR)\footnote{http://www.networkrevolution.co.uk/project-library/dataset-tc1a-basic-profiling-domestic-smart-meter-customers/} and UK power networks (UKPN)\footnote{https://data.london.gov.uk/dataset/photovoltaic--pv--solar-panel-energy-generation-data} respectively. The ESP's import  and export  prices are set to be \(\lambda_{b}^t = 0.05\) \pounds/kWh and \(\lambda_{s}^t = 0.03\) \pounds/kWh respectively.  

The prosumers P1, P2 and P3 all have 1 kW of installed solar capacity and are exposed to equal amounts of solar irradiance. Each prosumer has battery parameters\footnote{https://www.tesla.com/support/energy/powerwall/documents/documents/} as given in Table \ref{bat_params}. 
\begin{table}[h!]
	\centering
	\caption{Battery Parameters.}
	\begin{tabular}{l l}%			  
		\hline
		Parameter & Value/Description\\			
		\hline
		\hline		
		Battery type & Tesla Powerwall \\
		Life cycle & 5000\\ 
		Initial SoC & 50\%\\ 
		Usable capacity & 13.5 kWh\\			
		Depth of discharge & 100\%\\ 			
		Price per kWh (\pounds/kWh) & 314.64 \\		
		Round trip efficiency & 92.5\% \\				 
		\hline 
		\hline
	\end{tabular} 
	\label{bat_params}
\end{table}

The actor and critic networks for each prosumer agent are designed using hyper-parameters tabulated in Table \ref{hyper_params}. Algorithm \ref{algorithm} is developed and implemented in Python using PyTorch framework \cite{paszke2019pytorch}. An  OpenAI Gym environment \cite{brockman2016openai} is designed to model the multi-agent energy trading environment.
\begin{table}[h!]
		\centering
		\caption{Hyper-parameters for each Actor and Critic Network.}
		\begin{tabular}{l l l}%			  
			\hline
			Hyper-parameter& Actor Network & Critic Network\\			
			\hline
			\hline
			Optimizer & Adam & Adam \\
			Batch size & 256  & 256\\			
			Discount factor & 0.95  & 0.95\\ 			
			Learning rate & $1\times 10^{-4}$ & $3\times 10^{-4}$\\		
			No. of hidden layers & 2 & 2\\	
			No. of nodes in each layer & 500 & 500\\ 			 
			\hline 
			\hline
		\end{tabular} 
		\label{hyper_params}
\end{table}

\subsection{Performance Analysis Without Network Constraints}
\begin{figure}[!t]
	\centering
	\includegraphics[width=3.1in]{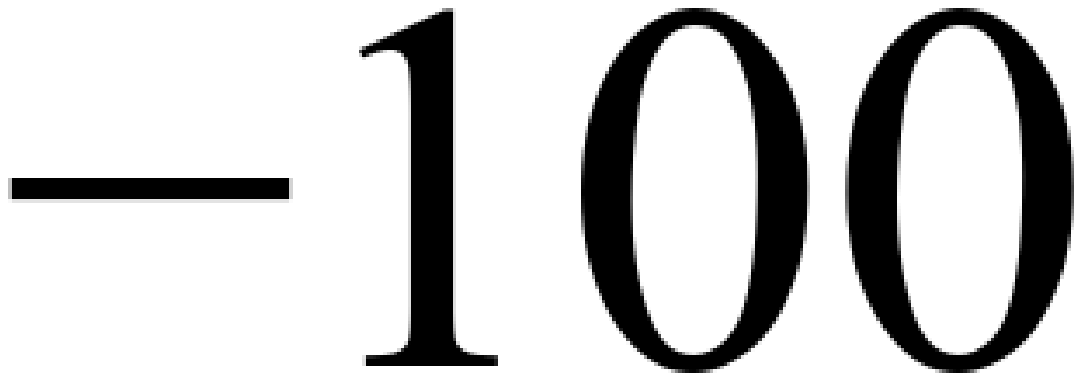}%
	\caption{Mean episode rewards of the agents during the training process.}
	\label{rewards1}
\end{figure}

\begin{figure}[!t]
	\centering
	\includegraphics[width=3.1 in]{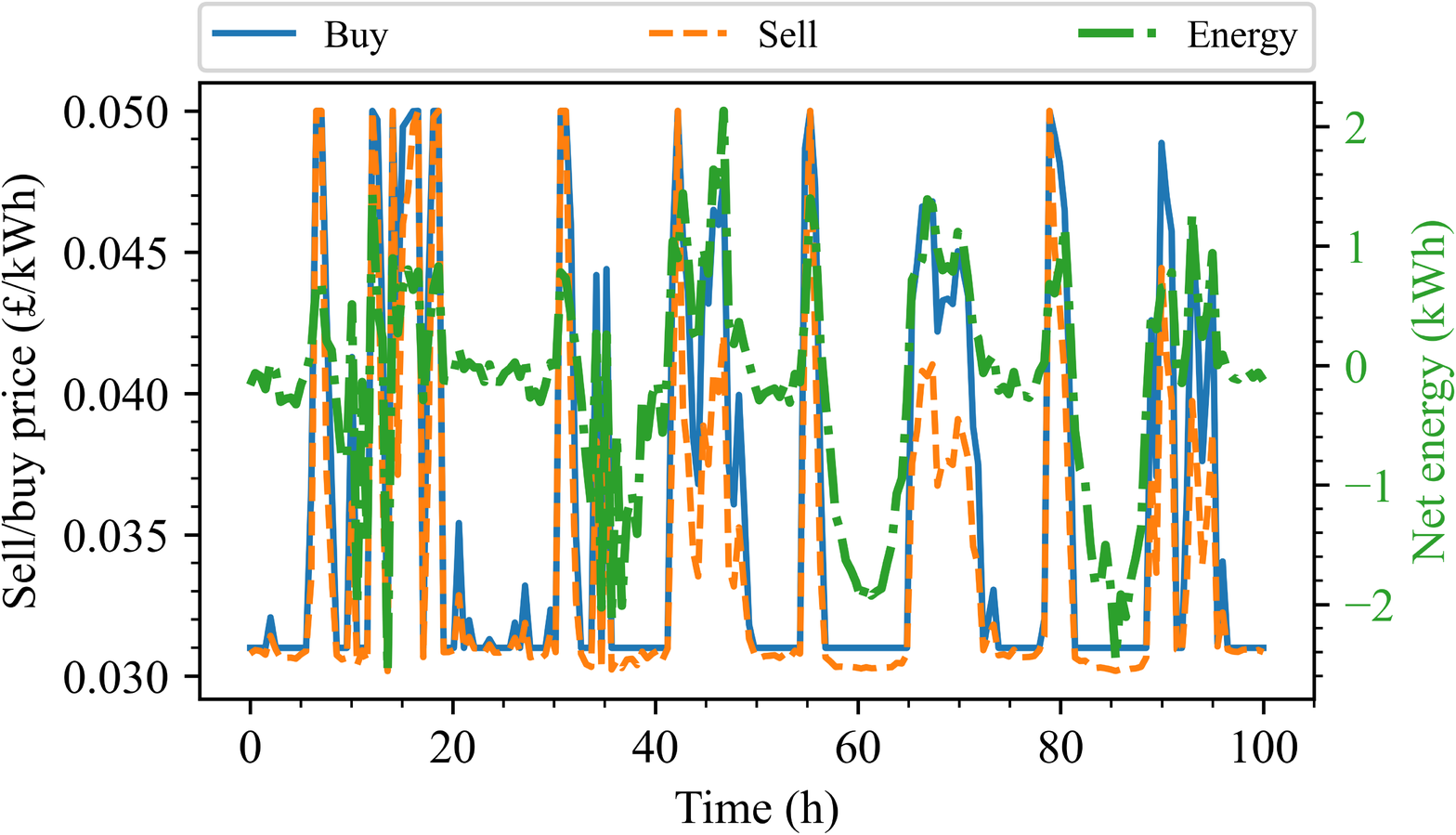}
	\caption{Variation of energy buying/selling price with total net energy of the prosumers without considering network constraints.}
	\label{pricevsnet1}
\end{figure}
In this section, convergence and performance analysis of the proposed algorithm without first considering distribution network constraints are presented. The agents are trained with 800 episodes and the evolution of the  episode rewards is shown in Fig. \ref{rewards1}. As the agents explore their action spaces according to the Ornstein-Uhlenbeck process
(\ref{action}), the episode rewards keep fluctuating until after 300 episodes when the training becomes stable. This shows that
the proposed MADDPG achieves stable trainings despite the energy trading environment being non-stationary from the perspective of each agent.

For presentation purposes, we show performance of the proposed algorithm using the first 100 hours of the dataset which is shown in Fig. \ref{profiles}. Fig. \ref{pricevsnet1} shows that the energy selling and buying price are high and low when net energy is positive and negative respectively. Positive net energy means that total consumption is more than total energy generation in the P2P energy sharing community, and the converse is true.

\begin{figure}[!t]
	\centering
	\includegraphics[width=3.1in]{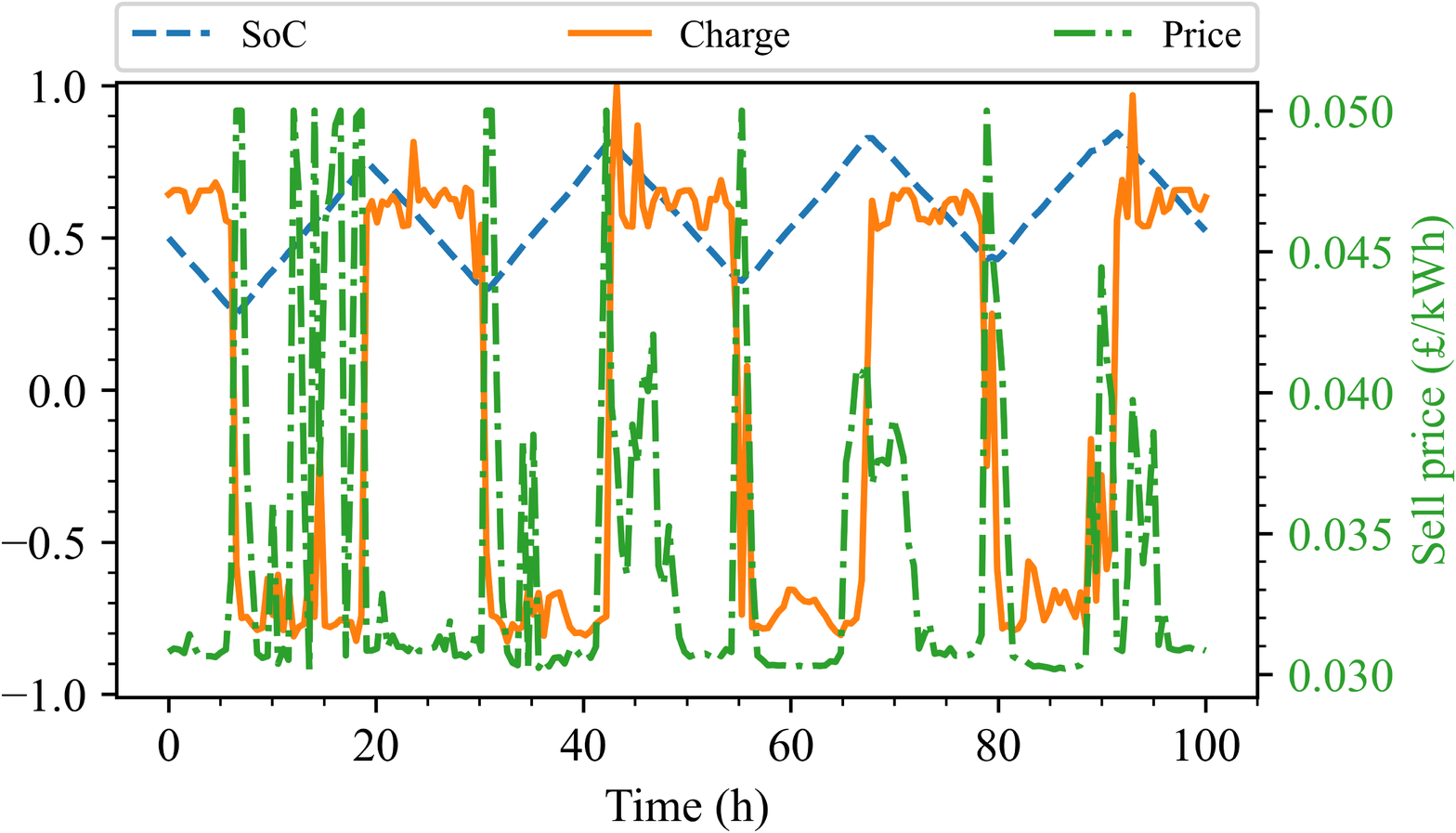}
	\caption{The charging (negative charge) and discharging (positive charge) action of the battery for prosumer 1 as it responds to the price (e.g. the selling price).}
	\label{scheduling}
\end{figure}
Using prosumer P1 as case study, Fig. \ref{scheduling} shows the optimal battery control actions. In the figure, the (negative) charge and (positive) discharge power of the battery are scaled to  -1 and 1 in order to plot on one graph the battery SoC and power output. We can observe that the proposed algorithm can learn the optimal policy to charge/discharge the battery optimally. That is, to charge the battery when the price (e.g. selling price) is low and discharge the battery when the price is high, thus, reducing the energy costs.

\subsection{Performance Analysis With Network Constraints}
\begin{figure}[!h]
	\centering
	\includegraphics[width=3.1 in]{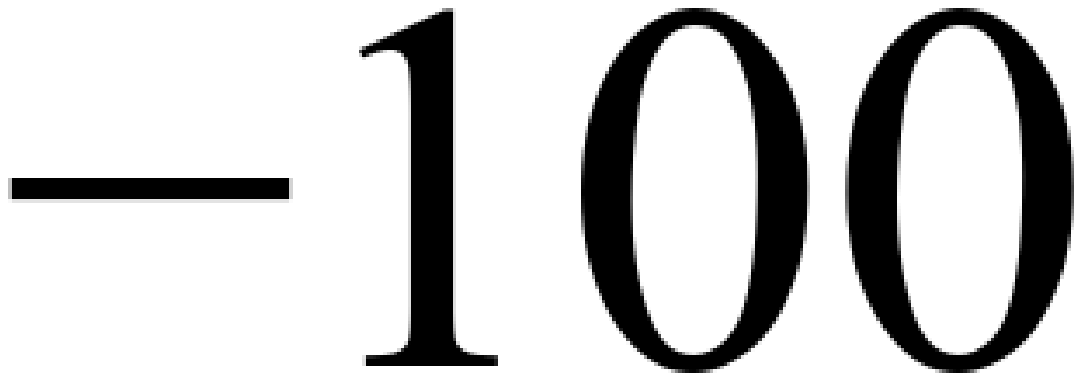}
	\caption{Mean episode rewards of the agents during the training while considering the network constraints.}
	\label{rewards2}
\end{figure}

\begin{figure}[!h]
	\centering
	\includegraphics[width=2.8 in]{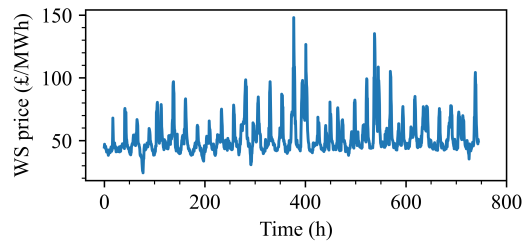}
	\caption{U.K's wholesale (WS) market price for the month of January, 2017.}
	\label{wholesale}
\end{figure}

In this section we evaluate the effectiveness of the proposed algorithm for charging and discharging the batteries optimally while satisfying distribution network constraints. We use U.K's January 2017 wholesale (WS) market electricity price  which is obtained from the institution of civil engineers (ICE)\footnote{https://www.ice.org.uk/knowledge-and-resources/briefing-sheet/the-changing-price-of-wholesale-uk-electricity/} as shown in Fig. \ref{wholesale} to determine the DNTs. We also assume that the loads have 0.95 power factor and the solar and battery energy storage systems have unity power factor. We present the results using the first 100 hours of the dataset in Fig. \ref{profiles}. Fig. \ref{rewards2} shows that learning is stable even when network constraints are considered.


\begin{table}[h!]
	\centering
	\caption{Average Episode Rewards.}
	\begin{tabular}{c l l c}%			  
		\hline
		Prosumer & Without DNTs & With DNTs & Difference (\%)\\			
		\hline
		\hline
		P1 & -2464.25 & -1919.88 & 22.1 \\
		P2 & -22661.05 & -1919.55 & 15.1\\			
		P3 & -2258.64 & -2073.27 & 8.2\\ 			 
		\hline 
		\hline
	\end{tabular} 
	\label{comp_rewards}
\end{table}

Table \ref{comp_rewards} compares the average  episode rewards of Fig. \ref{rewards1} and Fig. \ref{rewards2}. We can observe that prosumers benefit more (by having more than 8.2\% of accumulated episode rewards) when they support distribution network constraints than when they do not. Rewards are highest for P1 because it has the lowest total load: P1, P2 and P3 have a total load of 194 kWh, 219.17 kWh and  240.76 kWh respectively. Thus, much of the renewable energy generation is sold to other prosumers for profit, hence increasing the rewards. 

\begin{figure}[!t]
	\centering
	\subfloat[\label{dnt}]{\includegraphics[width=0.28\linewidth]{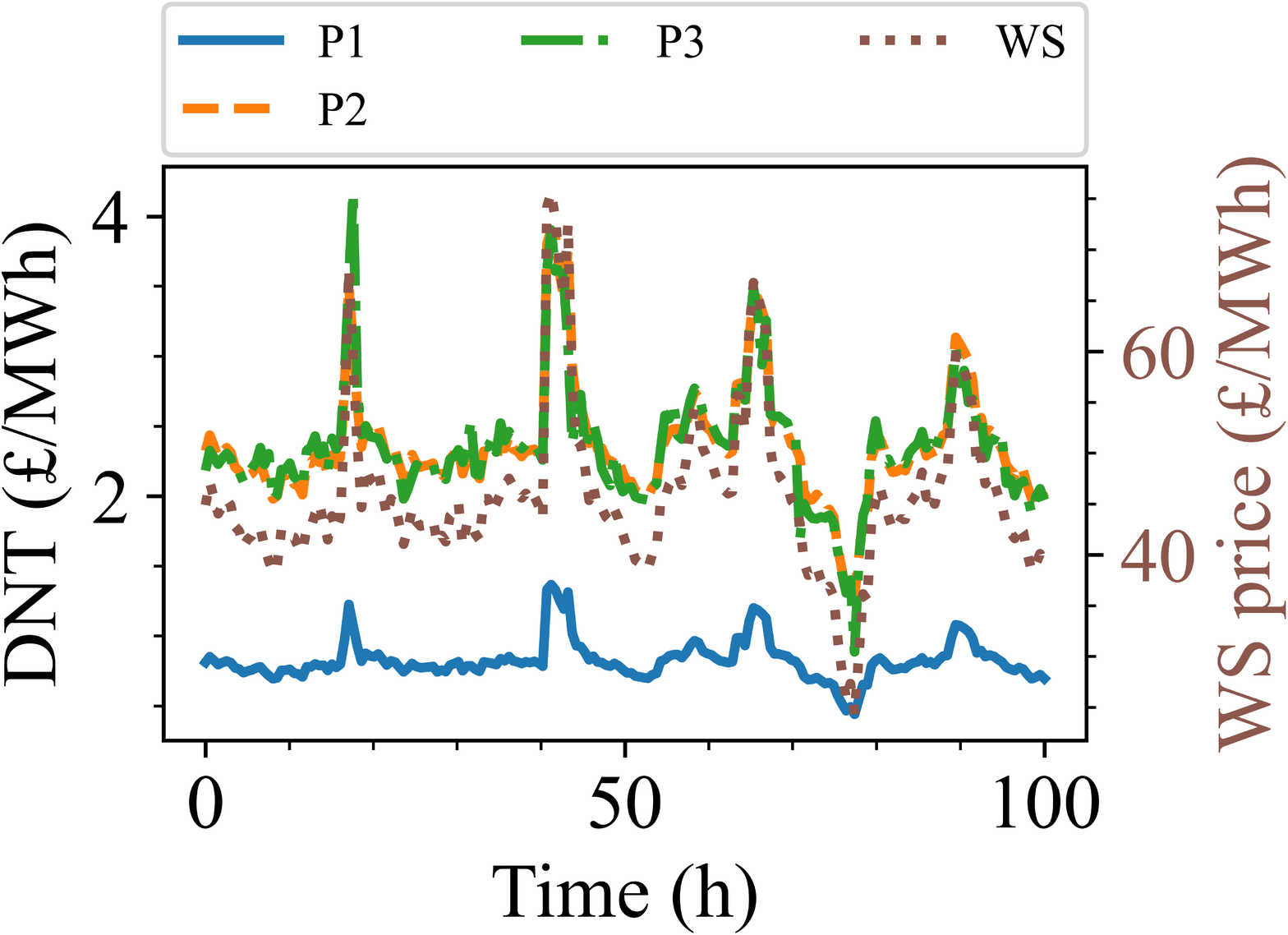}%
	}
	\hfil
	\subfloat[\label{loss_comp}]{\includegraphics[width=0.28\linewidth]{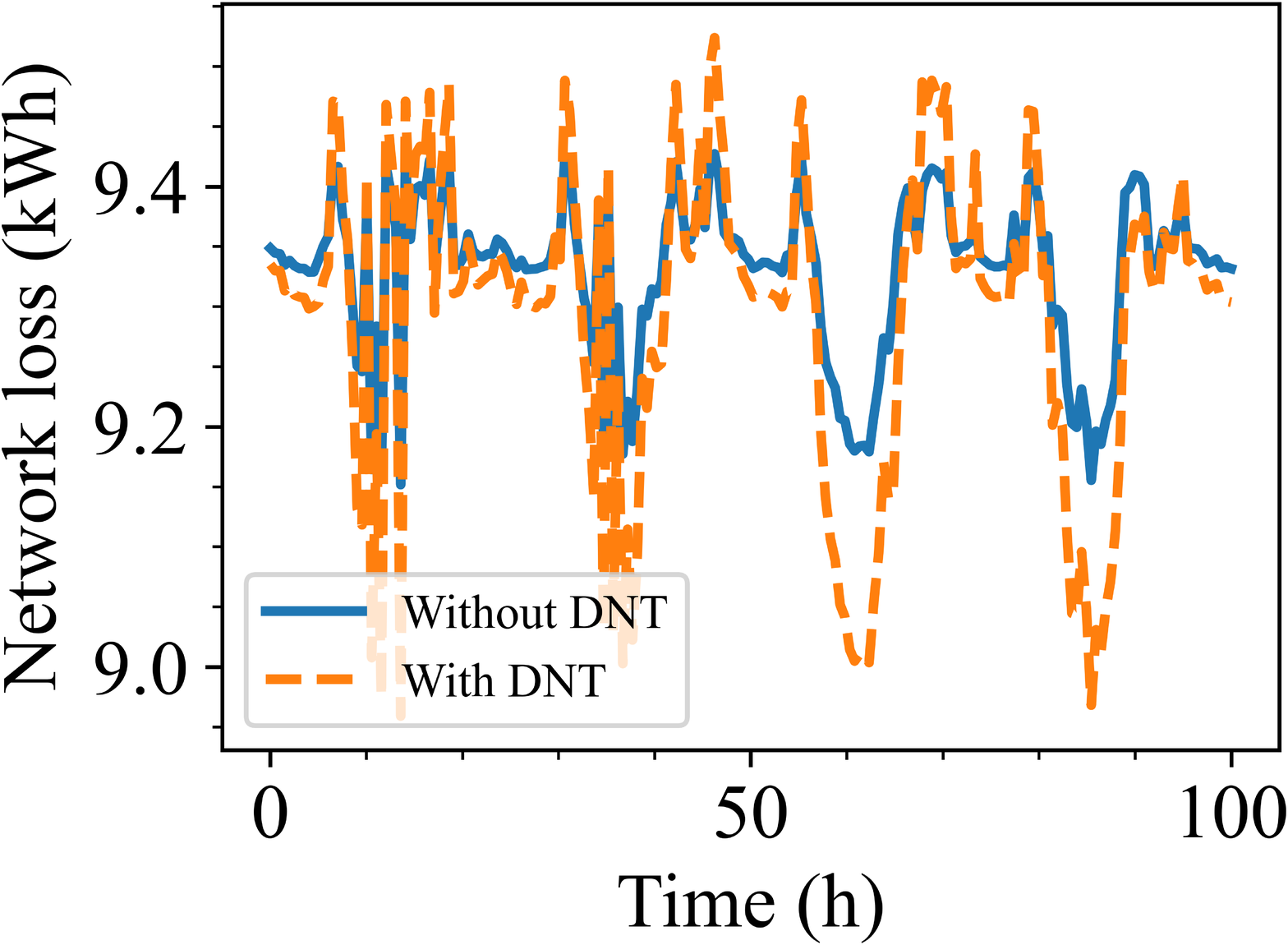}%
	}
	\caption{(a) Distribution network tariffs (DNT) for P1, P2 and P3, and (b) network loss comparison with and without DNTs.}
	\label{comp_loss}
\end{figure}

Further, Fig. \ref{dnt} shows that P1  contributes least to network loss, congestion and voltage limit violation as it has the lowest value of the DNT. The DNTs change according to the wholesale price, making them economically suitable for influencing the consumption pattern of prosumers. 

\begin{figure}[!t]
	\centering
	\subfloat[\label{voltage}]{\includegraphics[width=0.28\linewidth]{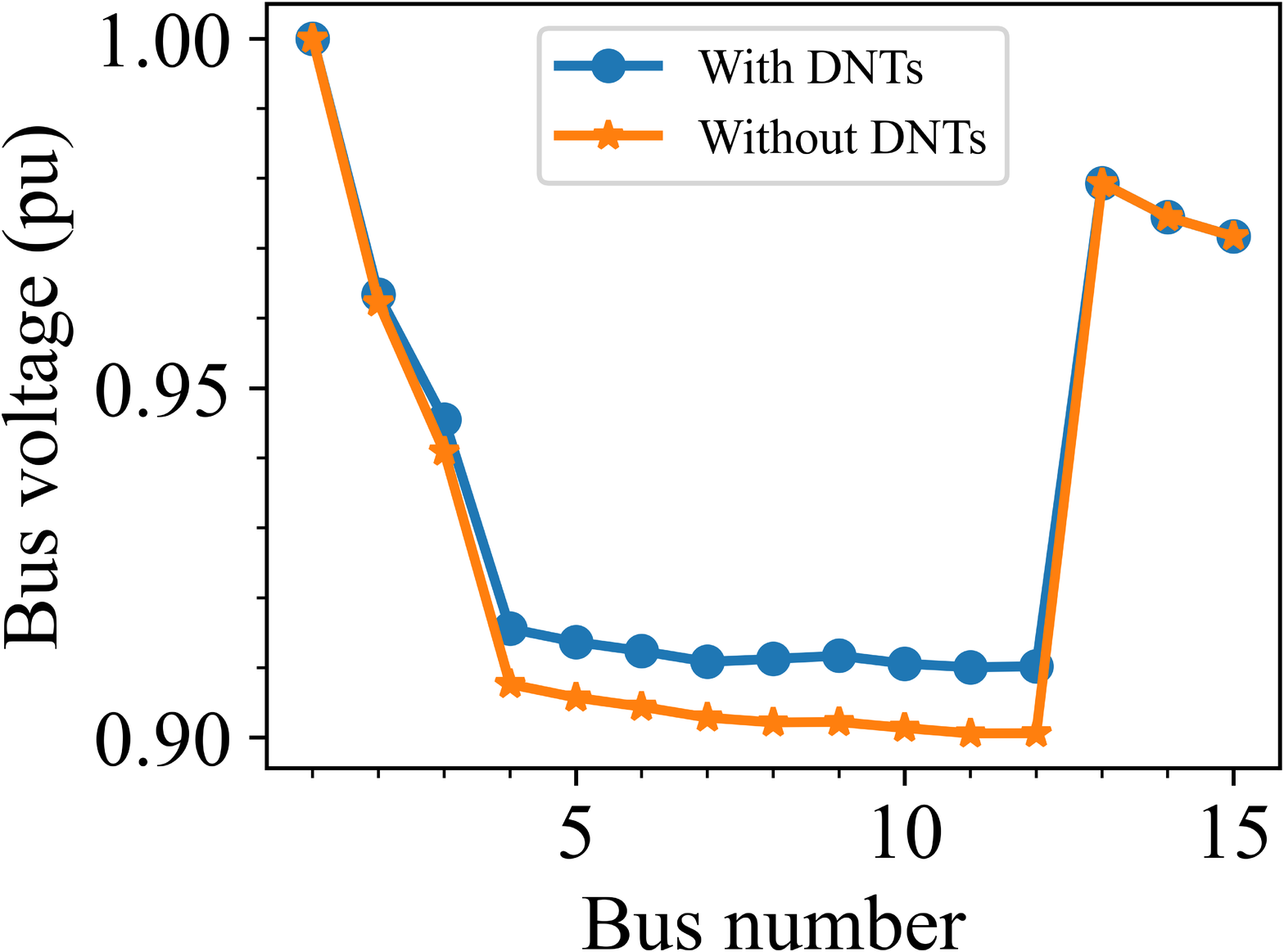}%
	}
	\hfil
	\subfloat[\label{line_loading}]{\includegraphics[width=0.28\linewidth]{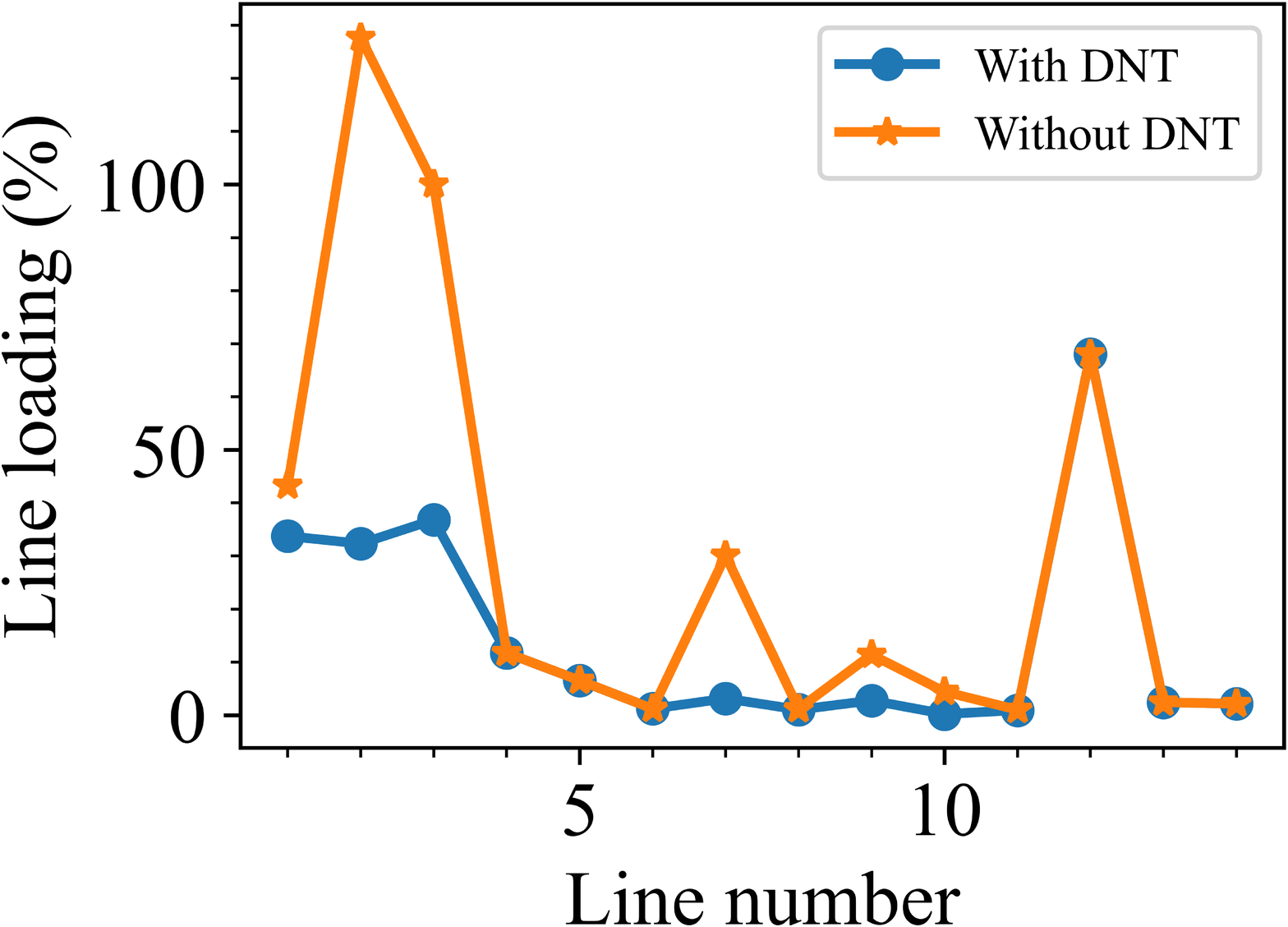}%
	}
	\caption{(a) Average bus voltage and (b) line loading with and without DNTs.}
	\label{comp_voltage_loading}
\end{figure}

Benefits of incorporating DNTs in the P2P energy selling and buying price to the distribution network are also shown in  Fig. \ref{loss_comp} and Fig. \ref{comp_voltage_loading}.  Fig. \ref{loss_comp} shows that  by incorporating the DNTs in the proposed algorithm, network losses are reduced. We can  observe in Fig. \ref{voltage} that voltage regulation is improved and in Fig. \ref{line_loading} that (peak) congestion is reduced by more than 50\% when DNTs are used. 

\subsection{Performance Comparison}
\begin{figure}[!t]
	\centering
	\includegraphics[width=3.5in]{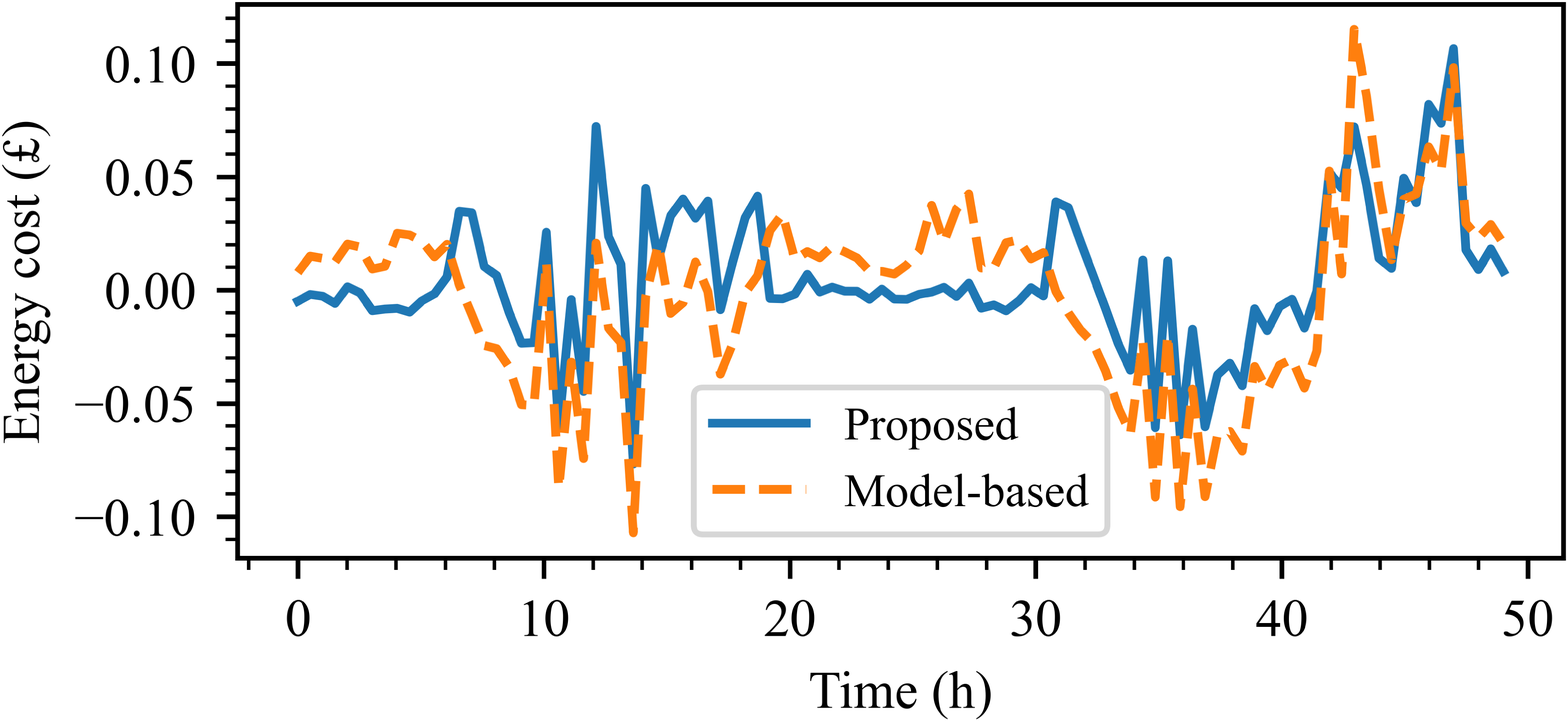}
	\caption{Comparison of total import (positive) and export (negative) energy costs for the prosumers between the proposed algorithm and  model-based approach.}
	\label{compare_cost}
\end{figure}

In this section, effectiveness of the proposed algorithm at reducing import energy costs from the main grid is compared to that of a model-based approach derived from the method detailed in  \cite{liu2017energy}.  To reduce complexity and the number of variables required by the model-based approach, the results are presented using the first 50 hours of the dataset in Fig. \ref{profiles}. The comparison result is shown in Fig. \ref{compare_cost}. We can observe that the energy cost result obtained by the proposed algorithm is consistent with that obtained from the model-based approach, verifying that the results produced by the proposed algorithm are accurate.

\section{Conclusion}\label{conc}
In this paper, we have proposed a MADDPG-based algorithm to minimize the energy costs of prosumers participating in peer-to-peer energy trading while considering the distribution network constraints. The energy costs are minimized by scheduling the operation of batteries optimally as flexible assets. First, the battery scheduling process is modelled as a Markov decision process. Then, the MADDPG algorithm proposed (which is model-free) is used to learn the optimal battery scheduling strategies that minimize the energy costs. To satisfy the distribution network constraints, we have proposed the use of DNTs. The DNTs act as incentives enticing the prosumers to either reduce or increase their consumption as a way of satisfying the network constraints. Simulation results based on real-world datasets have shown that the algorithm proposed can optimally minimize the energy costs while also satisfying the distribution network constraints. Minimizing the energy costs by also scheduling the operation of flexible loads is a potential future work.

\bibliographystyle{ieeetr}%{unsrtnat}
\bibliography{refs}  %%% Uncomment this line and comment out the ``thebibliography'' section below to use the external .bib file (using bibtex) .





% \end{thebibliography}

\end{document}